\documentclass[twocolumn,prb,showpacs,preprintnumbers,amsmath,amssymb]{revtex4}

\usepackage{graphicx}
\usepackage{booktabs}
\usepackage{bm}
\usepackage{topcapt}
\usepackage{booktabs}
\usepackage{longtable}
\usepackage{here}

\begin{document}


\title{Strongly three-dimensional electronic structure and Fermi Surfaces of SrFe$_{2}$(As$_{0.65}$P$_{0.35}$)$_{2}$: Comparison with BaFe$_{2}$(As$_{1-x}$P$_{x}$)$_{2}$}

\author{H. Suzuki$^{1}$, T. Kobayashi$^{2}$, S. Miyasaka$^{2,6}$, T. Yoshida$^{1,6}$, K. Okazaki$^{1}$, L. C. C. Ambolode II$^{1}$, S. Ideta$^1$, 
M. Yi$^3$, M. Hashimoto$^4$, D. H. Lu$^4$, Z.-X. Shen$^4$, 
 K. Ono$^5$, H. Kumigashira$^5$, S. Tajima$^{2,6}$ and A. Fujimori$^{1,6}$}
 
\affiliation{$^1$Department of Physics, University of Tokyo,
Bunkyo-ku, Tokyo 113-0033, Japan}

\affiliation{$^2$Department of Physics, Osaka University, Toyonaka, Osaka 560-8531, Japan}

\affiliation{$^3$Stanford Institute of Materials and Energy Sciences, Stanford University, Stanford, CA 94305, USA}

\affiliation{$^4$Stanford Synchrotron Radiation Lightsource, SLAC National Accelerator Laboratory, Menlo Park, CA 94305, USA}

\affiliation{$^5$KEK, Photon Factory, Tsukuba, Ibaraki 305-0801, Japan}

\affiliation{$^6$JST, Transformative Research-Project on Iron
Pnictides (TRIP), Chiyoda, Tokyo 102-0075, Japan}

\date{\today}

\begin{abstract}
The isovalent-substituted iron-pnictide superconductor SrFe$_{2}$(As$_{1-x}$P$_{x}$)$_{2}$ ($x$=0.35) has a slightly higher optimum critical temperature than the similar system BaFe$_{2}$(As$_{1-x}$P$_{x}$)$_{2}$, and its parent compound SrFe$_{2}$As$_{2}$ has a much higher N\'eel temperature than BaFe$_{2}$As$_{2}$. We have studied the band structure and the Fermi surfaces of optimally-doped SrFe$_{2}$(As$_{1-x}$P$_{x}$)$_{2}$ by angle-resolved photoemission spectroscopy (ARPES). Three holelike Fermi surfaces (FSs) around (0,0) and two electronlike FSs around ($\pi$,$\pi$) have been observed as in the case of BaFe$_{2}$(As$_{1-x}$P$_{x}$)$_{2}$. Measurements with different photon energies have revealed that the outermost hole FS is more strongly warped along the $k_{z}$ direction than the corresponding one in BaFe(As$_{1-x}$P$_{x}$)$_{2}$, and that the innermost one is an ellipsoidal pocket. The electron FSs are almost cylindrical unlike corrugated ones in BaFe(As$_{1-x}$P$_{x}$)$_{2}$. Comparison of the ARPES data with first-principles band-structure calculation revealed that the quasiparticle mass renormalization factors are different not only between bands of different orbital character but also between the hole and electron FSs of the same orbital character. By examining nesting conditions between the hole and electron FSs, we conclude that magnetic interactions between FeAs layers rather than FS nesting play an important role in stabilizing the antiferromagnetic order. The insensitivity of superconductivity to the FS nesting can be explained if only the $d_{xy}$ and/or $d_{xz/yz}$ orbitals are active in inducing superconductivity or if FS nesting is not important for superconductivity.
\end{abstract}


\maketitle
\section{INTRODUCTION}
Since the discovery of high temperature superconductivity in iron pnictides, the mechanism of Cooper pairing and the symmetry of the order parameter have been intensely debated. Among them the BaFe$_{2}$(As$_{1-x}$P$_{x}$)$_{2}$ (Ba122P) system \cite{Kasahara.S_etal.Phys.-Rev.-B2010} has attracted particular attention since the presence of line nodes in the superconducting (SC) order parameter was suggested experimentally \cite{Nakai.Y_etal.Phys.-Rev.-B2010,Hashimoto.K_etal.Phys.-Rev.-B2010,Yamashita.M_etal.Phys.-Rev.-B2011}. In the framework of the spin-fluctuation (SF)-mediated superconductivity mechanism, the intra-orbital nesting between hole and electron Fermi surfaces (FSs) brings about a sign-changing $s^{\pm}$-wave SC state \cite{Kuroki.K_etal.Phys.-Rev.-Lett.2008,Chubukov.AAnnual-Review-of-Condensed-Matter-Physics2012}. A suggested location is horizontal line nodes on the hole FS around the Z point \cite{Suzuki.K_etal.Journal-of-the-Physical-Society-of-Japan2011,Graser.S_etal.Phys.-Rev.-B2010}. On the other hand, in the framework of the orbital-fluctuation-mediated superconductivity mechanism, inter-orbital nesting between hole and electron FSs enhances antiferro-orbital fluctuation and a $s^{++}$-wave SC state without sign change will be realized \cite{Kontani.H_etal.Phys.-Rev.-Lett.2010,Onari.S_etal.Phys.-Rev.-Lett.2012,Saito.T_etal.Phys.-Rev.-B2010}. If both spin and orbital fluctuations are important, more complicated line nodes such as loop nodes have been predicted \cite{Saito.T_etal.Phys.-Rev.-B2013}. A loop-like node in the electron FS is also suggested based on a Raman scattering experiment \cite{Mazin.I_etal.Phys.-Rev.-B2010} and from the viewpoint of the hybridization between two electron pockets \cite{Khodas.M_etal.Phys.-Rev.-B2012}. In order to identify the pairing mechanism, systematic studies of the FS shapes and their orbital character are necessary. For this purpose, SrFe$_{2}$(As$_{1-x}$P$_{x}$)$_{2}$ (Sr122P), which is closely related to Ba122P, is an interesting system to study how the changes in the crystal structure affect the electronic, magnetic, and superconducting properties.  

\begin{figure}[htbp] 
  \centering
   \includegraphics[width=8cm]{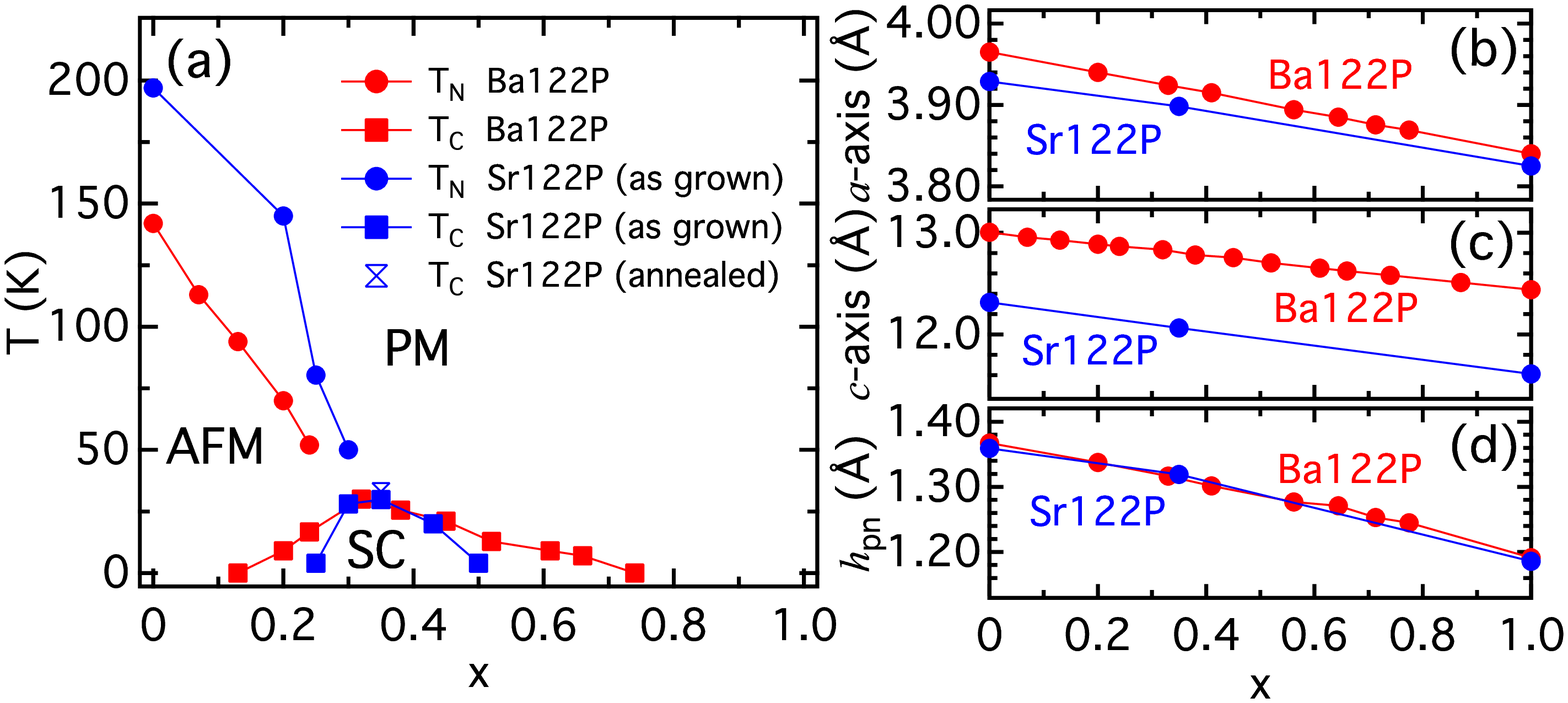} 
    \caption{(Color online) (a) Phase diagrams of BaFe$_{2}$(As$_{1-x}$P$_{x}$)$_{2}$ (Ba122P) \cite{Nakajima.M_etal.Journal-of-the-Physical-Society-of-Japan2012} and as-grown SrFe$_{2}$(As$_{1-x}$P$_{x}$)$_{2}$ (Sr122P) \cite{Kobayashi.T_etal.Journal-of-the-Physical-Society-of-Japan2012,Kobayashi.T_etal.Phys.-Rev.-B2013}. AFM, PM, SC stand for the antiferromagnetic, paramagnetic, and superconducting phases, respectively. (b), (c) Comparison of the $a$- and $c$-axis lattice constants. (d) Comparison of the pnictogen height $h_{\text{pn}}$. Data are taken from Refs. \onlinecite{Kobayashi.T_etal.Phys.-Rev.-B2013,Nakajima.M_etal.Journal-of-the-Physical-Society-of-Japan2012,Kasahara.S_etal.Phys.-Rev.-B2010,Saha.S_etal.Journal-of-Physics:-Conference-Series2011,Analytis.J_etal.Phys.-Rev.-Lett.2009}.}
   \label{phase}
\end{figure}

The phase diagrams of Ba122P and Sr122P are compared with each other in Fig. \ref{phase} (a) \cite{Nakajima.M_etal.Journal-of-the-Physical-Society-of-Japan2012,Kobayashi.T_etal.Journal-of-the-Physical-Society-of-Japan2012}. The N\'eel temperature ($T_{N}$) of the parent compound SrFe$_{2}$As$_{2}$ is 197 K, which is as much as 50 K higher than that of BaFe$_{2}$As$_{2}$, and $T_{N}$ is higher in Sr122P than in Ba122P in the doping range $x<0.3$, where antiferromagnetic (AFM) order is present. In the same way as Ba122P, superconductivity in Sr122P appears as the AFM order is suppressed by P substitution \cite{Dulguun.T_etal.Phys.-Rev.-B2012}. The superconducting transition temperature ($T_{c}$) of Sr122P reaches up to 30 K (as grown) or 33 K (annealed) at $x=0.35$ \cite{Kobayashi.T_etal.Phys.-Rev.-B2013}, which are comparable to, or higher than, $T_{c}=$ 30 K of BaFe$_{2}$(As$_{0.7}$P$_{0.3}$)$_{2}$. Superconductivity appears in the range $0.25<x<0.5$ for Sr122P, which is narrower than $0.13<x<0.74$ for Ba122P. 

The existence of line nodes is experimentally suggested in Sr122P as well. $^{31}$P-NMR and specific heat measurements show that Sr122P has a large residual DOS in as-grown samples below $T_{c}$ \cite{Dulguun.T_etal.Phys.-Rev.-B2012}. It was also found that annealing decreases the residual electronic specific heat coefficient at $T=0$, $\gamma_{r}$, and changes the magnetic field ($H$) dependence of $\gamma_{r}$ from sublinear $\gamma_{r} \propto H^{0.7}$ to $H$ linear $\gamma_{r} \propto H$ \cite{Kobayashi.T_etal.Phys.-Rev.-B2013}. These changes can be considered as a transition from a dirty to clean superconductor with reduced defect concentration.

The effect of replacing Ba atoms by smaller Sr atoms can be observed in the changes of the lattice parameters. Figures \ref{phase}(b)-(d) show the doping dependence of the $a$- and $c$-axis parameters, and the pnictogen height $h_{Pn}$, respectively \cite{Kobayashi.T_etal.Phys.-Rev.-B2013,Nakajima.M_etal.Journal-of-the-Physical-Society-of-Japan2012,Kasahara.S_etal.Phys.-Rev.-B2010,Saha.S_etal.Journal-of-Physics:-Conference-Series2011,Analytis.J_etal.Phys.-Rev.-Lett.2009}. All of them monotonically decrease as the doping level $x$ increases. Reflecting the smaller atomic radius of Sr atoms than Ba, the $a$-axis is slightly shorter and the $c$-axis is about  $ 0.8$ $\text{\AA}$ shorter in Sr122P. On the other hand, $h_{Pn}$, which is considered to be correlated with $T_{c}$ \cite{Kuroki.K_etal.Phys.-Rev.-B2009}, is almost the same in both systems. These results indicate that the main structural difference between Sr122P and Ba122P is the spacing between FeAs layers or the As-As distance in the $c$-direction, which may lead to different coupling strengths between the FeAs layers along along the $c$-axis.

In the present study, we have performed ARPES measurements of optimally-doped Sr122P ($T_{c}=30$ \text{K}, $x=0.35$, as grown) and compared the results with those of Ba122P \cite{Yoshida.T_etal.Phys.-Rev.-Lett.2011}  in order to elucidate the dependence of the band structure and FS shapes on the lattice parameters. ARPES data are compared with local-density-approximation (LDA) band-structure calculations in order to estimate the quasiparticle mass renormalization factors. We shall also discuss the importance of inter/intra-orbital nesting conditions from the obtained FS shapes and their orbital character.

\section{METHODS}

Sr122P ($x$=0.35) single crystals were prepared by the self-flux method described in Ref. \onlinecite{Kobayashi.T_etal.Phys.-Rev.-B2013}. $T_{c}=30$ K was determined from the onset of Meissner diamagnetic signal with a transition width $\Delta T_{c}\simeq$ 5 K. We have also measured annealed samples and observed the same band structures and FSs. Some bands were more clearly resolved in annealed samples \cite{supple}. ARPES experiments were carried out at beamline 28A of Photon Factory (PF), KEK, and beamline 5-4 of Stanford Synchrotron Radiation Lightsource (SSRL). In order to obtain fresh surfaces, all samples were cleaved \textit{in situ} at pressure better than $1\times10^{-10}$ Torr. Cleavage occurs along the $ab$ planes. The samples were cleaved and kept at $T=10$ K during the measurements. There was no sign of sample degradation during experiments of $\sim$ 24 hours. The kinetic energy and the momentum of photoelectrons were measured using Scienta SES2002 and R4000 electron energy analyzers  at PF and SSRL, respectively. In-plane ($k_{X}$,$k_{Y}$) and out-of-plane momenta ($k_{z}$) are expressed in units of $\pi/a$ and $2\pi/c$, where $a=3.90$ \AA \,and $c=12.09$ \AA\, are the in-plane and out-of-plane lattice constants. Here, the $X, Y$ axes point from Fe towards the second nearest neighbor Fe atoms and the $z$ axis is parallel to the $c$-axis. Calibration of the Fermi level ($E_{F}$) was achieved using spectra of gold which was in electrical contact with the samples. Incident photon energy from 24 eV to 88 eV were linearly polarized. The energy resolution was $\Delta E\sim$ 5 meV.

In order to study the orbital character of the observed bands and the effect of replacing Ba by Sr in Ba122P, we have performed LDA band-structure calculations using Wien2k package \cite{Blaha.P_etal.2001}. The $x$ and $y$ axes point from Fe towards the nearest neighbor Fe atoms. The calculations were done using the experimentally determined tetragonal lattice constants $a,c$ and the pnictogen heights $h_{pn}$. The lattice parameters used in the calculations (in units of \AA) are $a=3.92$, $c=12.76$, $h_{\text{pn}}=1.29$ for Ba122P and $a=3.90$, $c=12.09$, $h_{\text{pn}}=1.31$ for Sr122P. The calculations are done for the parent compounds Ba122 and Sr122 using the crystal structural parameters for $x=0.35$ because the band structures were almost identical for BaFe$_{2}$As$_{2}$ and BaFe$_{2}$P$_{2}$ if the same crystal structures were used.

\section{RESULTS AND DISCUSSION}
\begin{figure}[htbp] 
   \centering 
   \includegraphics[width=8cm]{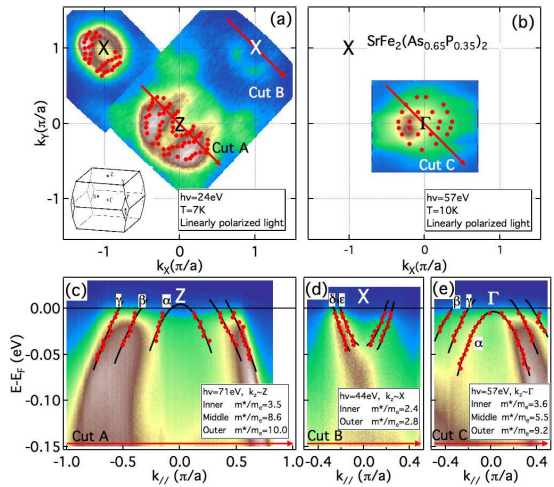} 
   \caption{(Color online) (a),(b) In-plane Fermi surface (FS) mapping for  Sr122P ($x=0.35$) taken at $h\nu=24$ eV [(a)] and $57$ eV [(b)]. Filled circles indicate the $k_{F}$ positions. The first Brillouin zone of Sr122P is shown in the inset of panel (a). (c)-(e) ARPES intensity plot taken along the cuts shown in (a) and (b). }
   \label{inplaneFS}
\end{figure}

 In-plane Fermi surface mapping is shown in Figs. \ref{inplaneFS} (a) and (b). Sr122P has a space group symmetry of I4/mmm and its first Brillouin zone is shown in the inset of (a). Changing the photoemission angle $\theta$ and $\phi$ with a fixed photon energy $h\nu$ corresponds to an approximately constant $k_{z}$ plane in the momentum space. There are three hole-like bands crossing $E_{F}$ around the Z point [Fig. \ref{inplaneFS} (c)], while two hole-like bands cross $E_{F}$ around the $\Gamma$ point [Fig. \ref{inplaneFS} (e)]. There are two electron-like bands crossing $E_{F}$ around the X point [Fig. \ref{inplaneFS} (d)]. 
  We shall discuss the orbital character of the FSs in the following paragraph. The difference of the intensities within the hole FSs around the Z point and the difference between the two identical electron pockets at $(1,1)$ and $(-1,1)$ are due to matrix-element effects arising from the geometrical setup of the experiment. 
 
 ARPES $E$-$k$ intensity plots around high-symmetry cuts are shown in Figs. \ref{inplaneFS} (c)-(e). By taking the second derivatives of the ARPES intensity, we have determined the peak positions of the multiple bands near $E_{F}$ (see Supplemental Material \cite{supple}). These peaks are fitted to parabolic dispersions of the form $E(\bm{k})=E(0)+\frac{\hbar^{2}k^{2}}{2m^{\ast}}$ and the quasiparticle mass $m^{\ast}/m_{e}$ is deduced. These values are shown in the inset of Figs. \ref{inplaneFS} (c)-(e). We shall discuss below the electron correlation strength by comparing the ARPES data with LDA band-structure calculation results.

\begin{figure}[htbp] 
   \centering
   \includegraphics[width=8cm]{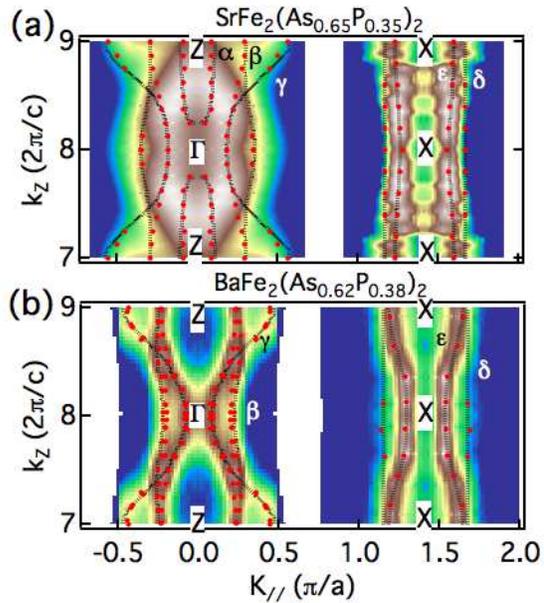} 
   \caption{(Color online) Fermi surface mapping taken in the $k_{z}$-$k_{\parallel}$ plane for Sr122P ($x=0.35$) [(a)] and Ba122P ($x=0.38$)\cite{Yoshida.T_etal.Phys.-Rev.-Lett.2011} [(b)]. In Sr122P, the $\gamma$ FS is warped more strongly than that of Ba122P and the $\alpha$ FS forms a 3D ellipsoidal hole pocket. The electron FSs $\epsilon$ and $\delta$ are less warped and nearly cylinder-like.}
   \label{kzFS}
\end{figure}

Figure \ref{kzFS} shows FS mapping along the $k_{z}$ direction taken with 30-90 eV photons. The relationship between the excitation energy and the $k_{z}$ position is given by the formula $k_z=\sqrt{\frac{2m}{\hbar^2}[(h\nu -\Phi)\cos^2 \theta +V_0]}$, where $h\nu$ is the photon energy, $\Phi$ is the work function, and $V_{0}$ is the inner potential \cite{Hufner.S_etal.2003}. $V_0$ is set to 13.5 eV for Ba122P and 14 eV for Sr122P in order to best reproduce the periodicity of the hole FS along the $k_{z}$ direction. The dotted curves in Fig. \ref{kzFS} represent the FS shapes determined by fitting $k_{F}$ positions. We call the inner, middle, and outer hole FSs around the Z point as $\alpha$, $\beta$, and $\gamma$ FSs, and the inner and outer electron FSs as $\epsilon$ and $\delta$ FSs, respectively. (The $\alpha$ FS of Ba122P was not visible in Ref. \onlinecite{Yoshida.T_etal.Phys.-Rev.-Lett.2011}.) 
In Sr122P, the $\gamma$ FS, which has strong $d_{z^{2}}$ character around the Z point, shrinks rapidly as it approaches the $\Gamma$ point. At the same time, the innermost $\alpha$ band shrinks and splits into two pockets. This is in contrast to that of Ba122P, in which all the hole FSs are connected in the entire $k_{z}$ region. The stronger three-dimensionality in Sr122P originates from enhanced interlayer hopping matrix elements due to the smaller $c$-axis lattice constant. 

The carrier number per Fe atom calculated from the FS volumes of Sr122P are $n_{\alpha}=0.041$, $n_{\beta}=0.64$, $n_{\gamma}=0.81$ (hole FSs), $n_{\delta}=0.39$ and $n_{\epsilon}=0.18$ (electron FSs), respectively. Although Sr122P is an isovalently substituted material, the total hole number $n_{h}=1.49$ estimated from ARPES is larger than the total electron number $n_{e}=0.57$.

\begin{figure}[htbp] 
   \centering
   \includegraphics[width=8cm]{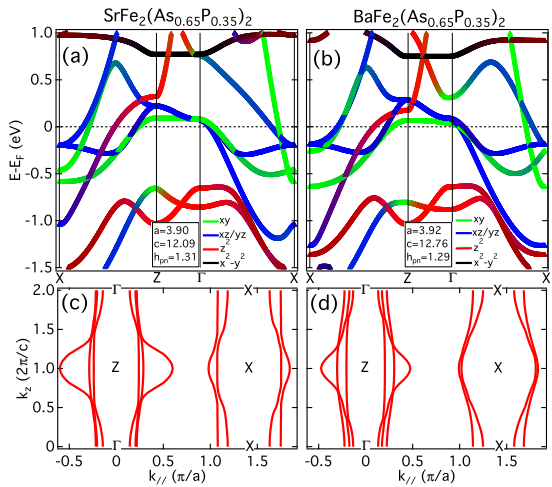} 
   \caption{(Color online) (a),(b) LDA band structures and their orbital character for Sr122P [(a)] and Ba122P [(b)] ($x$=0.35). The  lattice parameters are shown in the inset. (c),(d) Calculated FSs of Sr122P [(c)] and Ba122P [(d)]. }
   \label{Bandcomp}
\end{figure}

Figure \ref{Bandcomp} shows the LDA band structures and FSs of Sr122P and Ba122P ($x=0.35$). Around the $\Gamma$ point, the calculation predicts that two $d_{xz/yz}$ and one $d_{xy}$ hole bands cross $E_{F}$ in both systems. However, in the ARPES data of Sr122P [Fig. \ref{inplaneFS} (e)], the innermost band of $d_{xy}$ sinks slightly below $E_{F}$. The lowering of the $\alpha$ band by electron correlation probably causes this discrepancy.
Around the $Z$ point, three hole bands with $d_{xy}$ (inner), $d_{xz/yz}$ (middle), and $d_{z^{2}}$ (outer) orbital character, cross $E_{F}$, consistent with the ARPES data [Fig. \ref{inplaneFS} (c)]. The larger radius of the outer $d_{z^{2}}$ hole FS in Sr122P than that in Ba122P is also reproduced by the experiment. Around the $X$ point, the orbital characters of the two electron FSs ($d_{xy}, d_{xz/yz}$) along the $\Gamma$-$X$ line are interchanged between the two systems.

Comparison between the ARPES data and the LDA band-structure calculation enables us to evaluate the quasiparticle mass renormalization factor. Table \ref{massenhance} summarizes the orbital character, $m^{\ast}/m_{e}$, $m_{b}/m_{e}$, and $m^{\ast}/m_{b}$ of the bands near $E_{F}$, where $m^{\ast}$ is the effective mass, $m_{b}$ is the band mass, and $m_{e}$ is the free-electron mass. The masses listed here are the values estimated along the $\Gamma$-X line for the FSs around the $\Gamma$ and X points, and along the Z-X line for the FSs around the Z point by fitting the ARPES and LDA band to parabolic dispersions as shown in Fig. \ref{inplaneFS}. The $m^{\ast}/m_{b}$ values for Ba122P ($x=0.38$) are also shown as reference when direct comparison is possible\cite{supple2}. The $m^{\ast}/m_{e}$ values vary from 1.6 to 8.0 for Sr122P, and are larger than those of Ba122P, 1.2-5.5. The larger values of Sr122P may indicate stronger quantum critical fluctuations than in Ba122P \cite{Hashimoto.K_etal.Science2012}. A dynamical-mean-field-theory (DMFT) study\cite{Yin.Z_etal.Nat-Mater2011} predicts a positive correlation between the magnitude of the ordered magnetic moments and the mass renormalization of quasiparticles. The stronger renormalization in Sr122P with the higher $T_{N}$ is consistent with this trend.

Regarding the orbital dependence of mass enhancement, the DMFT calculation\cite{Yin.Z_etal.Nat-Mater2011} predicts that the enhancement is stronger for the $t_{2g}$ orbitals, $d_{xy}, d_{yz}$, and $d_{zx}$, than for the $e_{g}$ orbitals, $d_{z^{2}}$ and $d_{x^{2}-y^{2}}$, and is the strongest for $d_{xy}$. The strongest mass enhancement among the iron pnictides and chalcogenides, $m^{\ast}/m_{b}\sim 7$, takes place in the $d_{xy}$ orbital in FeTe, while the mass enhancement for $d_{xz/yz}$ in FeTe is $\sim$ 5. The orbital dependence of mass enhancement in Sr122P obtained here seems to follow this general trend, whereas the magnitudes are larger than the DMFT prediction in the $d_{xz/yz}$ orbital on the $\beta$ FS around the $\Gamma$ point, and in the $d_{xy}$ orbital on the $\epsilon$ FS around the X point. This difference indicates that each FS has different mass enhancement factors even for the same orbital character. Considering that the mass renormalization is expected to be enhanced by the scattering of quasiparticles between electron and hole FSs, the strong mass enhancement of the hole and electron FSs with different orbital character may imply the importance of inter-orbital scattering process.

\begin{table}[htbp]
   \centering
   \topcaption{Mass renormalization factors for the FSs of SrFe$_{2}$(As$_{0.65}$P$_{0.35}$)$_{2}$. $m^{\ast}$ is the effective mass, $m_{b}$ is the bare band mass, and $m_{e}$ is the free-electron mass. The values of $m^{\ast}/m_{e}$ and $m_{b}/m_{e}$ have been estimated by fitting the bands obtained from the ARPES (Fig. \ref{inplaneFS}) and the LDA calculations to parabolic dispersions in the range $\left|{E(\bm{k})}\right| <0.05$ eV. $m^{\ast}/m_{b}$ values for Ba122P ($x=0.38$) are also shown in parentheses when corresponding bands exist\cite{supple2}. Mass values for the anisotropic $\delta$ and $\epsilon$ FSs are calculated along the $\Gamma$-$X$ line.} 
   \begin{tabular}{lllccc} 
      \hline\hline
     $k_{z}$&FS&Orbital&$m^{\ast}/m_{e}$&$m_{b}/m_{e}$&$m^{\ast}/m_{b}$\\
       \hline
     $\Gamma$&$\beta$&$d_{xz/yz}$&9.2 (3.8) &1.6 (1.4) &5.8 (2.7)\\
   &$\gamma$&$d_{xz/yz}$&5.5 (2.6) &0.88 (0.72) &6.3 (3.6)\\
    Z&$\alpha$&$d_{xy}$&3.5&1.8&1.9\\
    &$\beta$&$d_{xz/yz}$&8.6 (2.5) &1.2 (1.2) &7.2 (2.0)\\
    &$\gamma$&$d_{z^{2}}$&10.0 (3.9) &2.7 (3.2) &3.7 (1.2)\\
    X&$\epsilon$&$d_{xy}$&2.4 (1.9) &0.30 (0.35) &8.0 (5.5)\\
    &$\delta$&$d_{xz/yz}$&2.8 (1.0) &1.8 (0.56) &1.6 (1.8)\\
      \hline\hline
   \end{tabular}
   \label{massenhance}
\end{table}

\begin{figure}[htbp] 
   \centering
   \includegraphics[width=8cm]{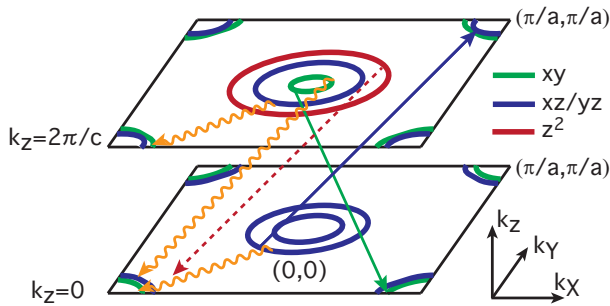} 
   \caption{(Color online) Schemetic presentation of the FS shapes and the orbital character on the $k_{z}\sim\, 0$ plane and the $k_{z}\sim \, 2\pi/c$ plane for Sr122P ($x=0.35$). The orbital character is determined by comparing the ARPES data and the LDA calculation. Antiferromagnetic fluctuations with wave vector $(\pi/a,\pi/a,2\pi/c)$ shown by solid arrows scatter quasi-particles between the electron and hole FSs of the same orbital character \cite{Kuroki.K_etal.Phys.-Rev.-B2009,Chubukov.AAnnual-Review-of-Condensed-Matter-Physics2012,Kemper.A_etal.New-Journal-of-Physics2010}. Since the $d_{z^{2}}$ band does not cross $E_{F}$ around the zone corner, it does not contribute to the pairing through the spin-fluctuation channel (dotted arrow). Wavy lines indicate nesting between FSs of different orbital character ($d_{xz/yz}$ and $d_{xy}$), which gives rise to orbital fluctuations and a finite SC gap on the $d_{z^{2}}$ FS \cite{Onari.S_etal.Phys.-Rev.-Lett.2012,Saito.T_etal.Phys.-Rev.-B2010,Kontani.H_etal.Phys.-Rev.-B2011,Shimojima.T_etal.Science2011}.}
   \label{FSGap}
\end{figure}

The FS shapes and their orbital character of Sr122P are summarized in Fig. \ref{FSGap}.
FSs of the same orbital character are connected by the antiferromagnetic ordering vector $\bm{Q}=(\pi/a,\pi/a,2\pi/c)$ \cite{Kuroki.K_etal.Phys.-Rev.-B2009,Chubukov.AAnnual-Review-of-Condensed-Matter-Physics2012,Kemper.A_etal.New-Journal-of-Physics2010}. Also, inter-orbital nesting between the $d_{xz/yz}$ and $d_{xy}$ orbitals, which enhances the antiferro-orbital fluctuation and brings about the large enhancement of the susceptibility in the charge channel \cite{Onari.S_etal.Phys.-Rev.-Lett.2012,Saito.T_etal.Phys.-Rev.-B2010,Kontani.H_etal.Phys.-Rev.-B2011}, is shown by wavy lines. 
On the $k_{z}\sim$ 0 plane, both hole FSs are of $d_{xz/yz}$ character and the $\alpha$ FS of $d_{xy}$ character is absent. On the $k_{z}\sim$ $2\pi/c$ plane [Fig. \ref{FSGap} (b)], the inner FS of $d_{xy}$ character is smaller than that in Ba122P ($x=0.30$) \cite{Shimojima.T_etal.Science2011} and the outer FS of $d_{z^{2}}$ character is larger due to the enhanced three-dimensionality. This modification in the FS radii deteriorate the intra-orbital nesting properties within the $d_{xy}$ orbital as compared with those in Ba122P. Besides, the effects of inter-orbital scattering seems to be also weaker in Sr122P, since the inter-orbital nesting properties between the enlarged $d_{z^{2}}$ FS and other FSs, and between the small ellipsoidal $d_{xy}$ FS and other FSs are worse. 

From this comparison we gain several insights into the origin of the AFM order and the superconductivity in the P-doped 122 systems. As for the AFM order, if the AFM order is described by the nesting-driven spin density wave, $T_{N}$ should be lower in Sr122P; however, $T_{N}$ is always higher in Sr122P as shown in Fig. \ref{phase} (a). We, therefore, suggest the importance of magnetic interaction between FeAs layers to explain the high $T_{N}$ of Sr122P with the shorter $c$-axis length. A mechanism of interlayer hybridization due to Fe-As-As-Fe hopping has indeed been suggested by Khodas \textit{et al.} \cite{Khodas.M_etal.Phys.-Rev.-B2012}. As for the superconductivity, although both inter- and intra-orbital nesting condition for the $\gamma$ hole FS is significantly worse in Sr122P than in Ba122P, $T_{c}=33$ K for annealed Sr122P samples is slightly higher than Ba122P; this can be explained if the $d_{xy}$ and/or $d_{xz/yz}$ are active while the $d_{z^{2}}$ orbital is not inducing superconductivity or if FS nesting is not important for superconductivity.

Finally, we comment on the suppression rates of $T_{N}$ and $T_{c}$ by P substitution\cite{Ishida.S_etal.Journal-of-the-American-Chemical-Society2013}. In the underdoped region ($x<0.3$), $T_{N}$ is more rapidly suppressed in Sr122P ($-dT_{N}/dx\sim 480 \text{K}/x$) than in Ba122P ($\sim 360 \text{K}/x$). Also, in the overdoped region ($x>0.3$), $T_{c}$ is more rapidly suppressed in Sr122P ($-dT_{c}/dx\sim 94 \text{K}/x$) than in Ba122P ($\sim 75 \text{K}/x$). These results indicate that the random potential induced by P substitution is stronger in Sr122P  than in Ba122P. This is probably because substituted P atoms in the more closely spaced FeAs layers [see Fig. \ref{phase} (c)] disturb the potential in the FeAs layers more strongly. In spite of the stronger impurity potential, $T_{c}$ is higher in Sr122P; this may be because the FS shapes with poorer nesting are more favorable to superconductivity, but further studies are necessary to clarify this issue.
 
\section{CONCLUSION}
We have performed ARPES measurements on the isovalent-substituted iron pnictide superconductor Sr122P  and compared the results with those of Ba122P. The outer hole FS of $d_{z^{2}}$ character is strongly warped along the $k_{z}$ direction than that of Ba122P. The dramatic change of the FS shapes induced by the replacement of Ba by Sr highlights the sensitiveness of the electronic structure on the $c$-axis parameter. Comparison of ARPES data with LDA band-structure calculations has revealed that quasiparticle mass is strongly renormalized and that the renormalization factors on the electron and hole FSs are different even within the same orbital. This may indicate that quasi-particles are scattered between electron and hole FSs of different orbital character. We have also examined nesting conditions between the hole FSs at the zone center and the electron FSs at the zone corner. The high $T_{N}$ of Sr122P in spite of the weaker intra/inter-orbital nesting indicates that the interlayer magnetic interactions play a more important role than FS nesting in stabilizing the AFM order. Also, despite the significantly worse nesting of the $d_{z^{2}}$ FS in Sr122P, $T_{c}=33$ K is slightly higher than that of Ba122P. This can be explained if only $d_{xy}$ and/or $d_{xz/yz}$ orbitals participate in superconductivity or if FS nesting is not important for superconductivity. These results impose restriction on microscopic theories and requires them to naturally reproduce the high $T_c$ of Ba/Sr122P and iron pnictides in general.

We are grateful to M. Nakajima and H. Eisaki for enlightening discussions and for providing phase diagram data of Ba122P. The
Stanford Synchrotron Radiation Lightsource is operated by the Office of Basic Energy Science,
US Department of Energy. Experiment at Photon Factory was approved by the Photon Factory Program Advisory
Committee (Proposal No. 2009S2-005 and 2012G075). H.S. acknowledges financial support from Advanced Leading Graduate Course for Photon Science (ALPS).

\bibliography{SuzukiSr}

\end{document}